\documentclass{article}

\usepackage{arxiv}

\usepackage[utf8]{inputenc} 
\usepackage[T1]{fontenc}    
\usepackage{hyperref}       
\usepackage{url}            
\usepackage{booktabs}       
\usepackage{amsfonts}       
\usepackage{nicefrac}       
\usepackage{microtype}      
\usepackage{lipsum}		
\usepackage{graphicx}
\usepackage{natbib}
\usepackage{doi}
\usepackage{graphicx}
\usepackage[normalem]{ulem}
\usepackage{multirow}%
\usepackage{amsmath,amssymb,amsfonts}%
\usepackage{array}
\usepackage{mathrsfs}%
\usepackage[title]{appendix}%
\usepackage{xcolor}%
\usepackage{textcomp}%
\usepackage{manyfoot}%
\usepackage{booktabs}%
\usepackage{algorithm}%
\usepackage{algorithmicx}%
\usepackage{algpseudocode}%
\usepackage{listings}%
\usepackage{subcaption}
\usepackage{placeins}
\usepackage{svg}

\begin{document}
\title{Learning Code-Edit Embedding \\ to Model Student Debugging Behavior}
\renewcommand{\shorttitle}{Learning Code-Edit Embedding to Model Student Debugging Behavior}

\author{{Hasnain Heickal}\\
	University of Massachusetts Amherst\\
    Amherst, MA 01002\\
	\texttt{hheickal@cs.umass.edu} \\
	\And
	{Andrew Lan}\\
	University of Massachusetts Amherst\\
    Amherst, MA 01002\\
	\texttt{andrewlan@cs.umass.edu} \\
}

%
\maketitle              
\begin{abstract}
Providing effective feedback for programming assignments in computer science education can be challenging: students solve problems by iteratively submitting code, executing it, and using limited feedback from the compiler or the auto-grader to debug. Analyzing student debugging behavior in this process may reveal important insights into their knowledge and inform better personalized support tools. In this work, we propose an encoder-decoder-based model that learns meaningful code-edit embeddings between consecutive student code submissions, to capture their debugging behavior. Our model leverages information on whether a student code submission passes each test case to fine-tune large language models (LLMs) to learn code editing representations. It enables personalized next-step code suggestions that maintain the student's coding style while improving test case correctness. Our model also enables us to analyze student code-editing patterns to uncover common student errors and debugging behaviors, using clustering techniques. Experimental results on a real-world student code submission dataset demonstrate that our model excels at code reconstruction and personalized code suggestion while revealing interesting patterns in student debugging behavior. \footnote{The code for the paper can found in: \url{https://github.com/umass-ml4ed/code-edit-representation}} \footnote{The paper is accepted in 26th International Conference
on Artificial Intelligence in Education (AIED 2025)}
\end{abstract}

\section{Introduction}\label{sec:intro}
Feedback is one of the most effective mechanisms for human learning since it helps learners reflect on their errors and refine their understanding through iteration \cite{hattie2007power}. However, delivering timely and personalized feedback is crucial for student learning \cite{Johnston2015TheEO,Kochmar2020AutomatedPF}, yet it remains challenging due to large class sizes and limited instructional resources \cite{narciss2008feedback}. When feedback is provided, it is often delayed until the final submission of an assignment, leaving students with little opportunity to act on it and correct their errors. In particular, for programming assignments in computer science, this issue is more evident, since students usually iterate between submitting code, executing it, receiving compiler feedback, and submitting an edited code. This iterative process plays a crucial role in developing their problem-solving and debugging skills \cite{keuning2020student}.  

In practice, many programming instructors rely on auto-grading tools such as Gradescope \cite{gradescope} and CodePost \cite{codepost}, which enable students to submit their code and receive feedback on whether their code passes a series of test cases \cite{singh2021building}, an implicit feedback that hints at what bugs they may have. However, this feedback is not personalized and tailored to the specific errors in a student's code, resulting in several limitations: First, if the system does not reveal the test cases, students may struggle to identify and correct their errors, particularly for those who lack strong debugging skills \cite{watson2014failure}. Second, while instructors can configure pre-written feedback messages for test case failures, this approach lacks personalization and may not effectively address individual student needs. Third, providing too much information, such as full test case inputs and expected outputs, can diminish learning; doing so reduces the cognitive effort students must undertake to diagnose and fix their errors. The concept of ``desirable difficulty'' suggests that learning is enhanced when students must engage with retrieval and problem-solving, rather than simply following explicit instructions \cite{bjork2011making}.  

An underutilized resource in programming education is the vast database of past student submissions, as programming assignments are often reused across semesters. A promising approach to leveraging these past submissions is to identify similar struggle patterns among students and suggest meaningful next steps based on prior solutions. Several works have explored different techniques for next-step hint generation. \cite{paassen2017continuous} formulated a mathematical framework for edit-based hint generation policies, providing next-step hints in the edit-distance space. \cite{piech2015learning} introduced a neural network-based method that encodes programs as a linear mapping from an embedded pre-condition space to an embedded post-condition space, enabling scalable feedback generation. More recent works have explored the use of large language models (LLMs) for automated feedback. \cite{phung2023hintgen} investigated the role of LLMs in providing human tutor-style programming hints. \cite{pankiewicz2023feedback} conducted an experimental study using GPT-3.5 to generate real-time feedback, measuring its impact on student learning. While these approaches introduce automation into feedback generation, they lack personalization, which has been shown to enhance student learning gains.  

A step toward personalized feedback is seen in \cite{heickal2024generating}, which introduces feedback at varying levels of detail, allowing for minor personalization. However, their study highlights a common issue with LLM-based approaches: instead of generating constructive hints, models like GPT-4 often provide direct solutions, which can diminish learning opportunities. Additionally, clustering student submissions to reveal common errors can help instructors identify recurring debugging challenges and improve course material accordingly \cite{emerson2020cluster}.  

While these works contribute to understanding student code and providing hints, they primarily focus on generating feedback based on a single code submission rather than analyzing the progression of a student’s code across multiple submissions. Debugging is an iterative process, and understanding how students modify their code over time—by examining consecutive submissions—can provide deeper insights into common debugging strategies and struggles. Existing work has not systematically explored how students refine their solutions step by step or how past debugging behaviors can inform personalized recommendations.  

Given these gaps, we explore the following research questions:  
\begin{itemize}
    \item \textbf{RQ1:} Can we utilize the edit patterns from historical submissions to generate personalized next-step code suggestions for students? 
    \item \textbf{RQ2:} Can we analyze these edit patterns to uncover common errors and typical debugging behaviors?  
\end{itemize}

In this work, we propose an encoder-decoder-based model that learns explicit vector representations of code-edits. Our model optimizes a contrastive loss function to structure the edit embedding space such that semantically similar edits are placed closer together. To guide this objective, we leverage test case transition patterns, defining two edits as similar if they lead to the same test case outcomes. Additionally, the model uses a code reconstruction objective to ensure that edit embeddings can generate meaningful next-step code suggestions. It also uses a regularization loss to maintain consistency in the embedding space, aligning edit vectors with the code embedding space.  

We run experiments on a real-world student code submission dataset and compare our model with GPT-4o. We find that our model generates edit embeddings that enable more personalized and meaningful next-step code suggestions. While GPT-4o often generates fully correct solutions, which may reduce student engagement and learning, our model generates intermediate code suggestions that try to focus on one error at a time: it passes a subset of previously failed test cases, while also mimicking the original student code submission as much as possible. We also analyze clusters of code-edit embeddings to identify common errors and debugging patterns among students.  

\section{Methodology}\label{sec:method}
\begin{figure}[t]
    \centering
    \includegraphics[width=\textwidth]{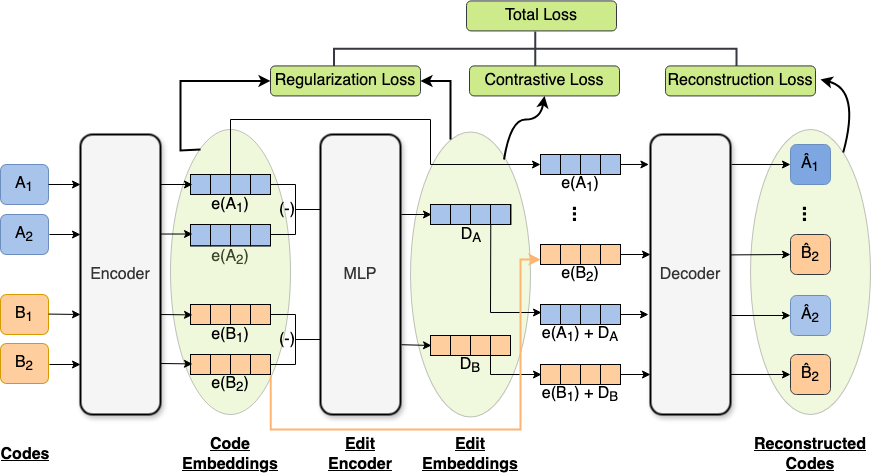}
    \caption{Our model architecture to learn code-edit embeddings.}
    \label{fig:model-arch}
\end{figure}
In this section, we describe our approach for learning meaningful code-edit representations to model student debugging behavior. We first discuss the notations and provide an overview of our model architecture. Next, we outline the training objectives: contrastive learning for edit representation, a decoding objective for reconstructing code, and a regularization objective to align embedding spaces. 

\subsection{Notations}
To learn code-edit representations, we investigate groups of four code submissions that we denote as $A_1, A_2, B_1, B_2$. Here, $C_A = (A_1, A_2)$ is a \emph{code-edit pair}: it corresponds to two consecutive code submissions to a problem $P$, by student $S_A$. Similarly, $C_B = (B_1, B_2)$ corresponds to two consecutive submissions to the same problem $P$ by (possibly a different) student $S_B$. Then we assign a label $Y_{A,B} \in \{0,1\}$ to this combination of four code submissions, where 1 means that $C_A$ and $C_B$ are similar code edits, and 0 otherwise. 

To define similar code edits, we resort to \emph{test cases}: our intuition is that if two code-edit pairs have the same test case passing/failing status, both before and after the edits, then they represent similar code-editing behavior. Specifically, assume that $A_1$ passes a subset of the test cases, $T_A^1$, and $A_2$ passes a subset of test cases, $T_A^2$. Similarly, for the code-edit pair $C_B$, we have $B_1$ and $B_2$ pass test case subsets $T_B^1$ and $T_B^2$, respectively. We have $Y_{A,B} = 1$ if both $T_A^1 = T_B^1$ and $T_A^2 = T_B^2$. This way, we obtain a binary similarity metric across all code-edit pairs. Since whether a code submission passes a test case is binary-valued, we refer to $T_A^1, T_A^2, T_B^1, T_B^2$ as the \emph{test-case-masks} of the corresponding codes.

\subsection{End-to-End Encoder-Decoder Model}

We use CodeT5 encoder-decoder model \cite{wang2021codet5} to learn code-edit representations. CodeT5 is well-suited for code-related tasks and offers a balance between performance and computational efficiency, making it more practical than larger models like CodeLLaMA or StarCoder, which require significantly more resources. Its encoder-decoder architecture allows structured learning of code edits, aligning well with our contrastive and decoding objectives. We fine-tune CodeT5 on student codes using a contrastive objective to learn meaningful code-change representations, a decoding objective to reconstruct code from embeddings, and a regularization objective to align the code and edit embedding spaces.

The model architecture is shown in Figure \ref{fig:model-arch}. It consists of an encoder, a decoder, and a layer for computing code-edit embeddings. The encoder processes code submissions into code token sequences and generates fixed-size embeddings by averaging the last hidden layer outputs for each code token, denoted as \( \mathbf{e}(A_1), \mathbf{e}(A_2), \mathbf{e}(B_1), \mathbf{e}(B_2) \in \mathbb{R}^{d} \), where \( A_1, A_2, B_1, B_2 \) are code submissions, and \( d \) denotes the dimension of the code embeddings. To capture edits, we compute code-edit embeddings for each pair \( (A_1, A_2) \) and \( (B_1, B_2) \) using a transformation parameterized by a fully-connected multi-layer perceptron neural network:  

\[
\mathbf{D}_A = \mathbf{W}_2 (\tanh(\mathbf{W}_1 (\mathbf{e}(A_2) - \mathbf{e}(A_1)) + \mathbf{b}_1)) + \mathbf{b}_2,
\]
\[
\mathbf{D}_B = \mathbf{W}_2 (\tanh(\mathbf{W}_1 (\mathbf{e}(B_2) - \mathbf{e}(B_1)) + \mathbf{b}_1)) + \mathbf{b}_2.
\]

Here, \( \mathbf{W}_1, \mathbf{W}_2 \in \mathbb{R}^{d \times d} \) and biases \( \mathbf{b}_1, \mathbf{b}_2 \in \mathbb{R}^{d} \) are learnable parameters. This structure enables the model to capture complex transformations between code versions, instead of assuming that the code embeddings and our designed code-edit embeddings are naturally aligned in the same vector space. 
The decoder, another CodeT5 model, reconstructs the original code token sequences from embeddings and generates transformed sequences using the edit representations. It operates in two modes: (i) \textbf{Direct Reconstruction}, where the decoder takes \( \mathbf{e}(A_1) \) and reconstructs \( A_1 \), and similarly for $A2, B1, B2$; (ii) \textbf{Edit Transformation}, where the decoder takes \( \mathbf{e}(A_1) + \mathbf{D}_A \) to generate \( A_2 \) and \( \mathbf{e}(B_1) + \mathbf{D}_B \) to generate \( B_2 \).  


The training process involves multiple loss functions. To ensure that the computed edit representations \( \mathbf{D}_A \) and \( \mathbf{D}_B \) capture meaningful code differences, we introduce a contrastive loss objective:  
\begin{align*}
d = \|\mathbf{D}_A - \mathbf{D}_B\|_2, \quad 
\mathcal{L}_{\text{C}} = Y_{A,B} \cdot d^2 + (1 - Y_{A,B}) \cdot \max(0, m - d)^2,
\end{align*}
where \( m \) is a margin hyperparameter for dissimilar pairs, and \( \| \cdot \|_2 \) denotes the Euclidean norm. Intuitively, this objective encourages code-edit embeddings $\mathbf{D}_A$ and $\mathbf{D}_B$ to be close to each other if their pre- and post-edit code submissions pass the same set of test cases, and pushes them farther away otherwise. Additionally, we introduce a reconstruction loss that ensures that the decoder's output reconstructs the actual student code submission from its embedding:
\[
\mathcal{L}_{\text{Rec}} = -\sum_{t=1}^T \log P([A_1]_t | [A_1]_{<t}, \mathbf{e}(A_1)),
\]
where \( [A_1]_t \) is the \( t \)-th token in the student code submission $A_1$ and \( [A_1]_{<t} \) represents all preceding tokens. Note that this objective applies to other code submissions $A_2$, $B_1$, and $B_2$ as well, which we omit for brevity. To further ensure that the learned code-edit embeddings are aligned with the original code embedding space, we introduce another regularization loss:  
\[
\mathcal{L}_{\text{Reg}} = \text{MSE}\left(A_1 + D_A, A_2\right) = \left\|\mathbf{e}(\mathbf{A}_1) + \mathbf{D}_A - \mathbf{e}(\mathbf{A}_2) \right\|_2^2,
\]
where \( \text{MSE} \) is the Mean Squared Error function. This loss term applies to code submissions $B_1$ and $B_2$ as well, which we again omit for brevity. 
The overall training objective is a weighted sum of the contrastive, reconstruction, and regularization losses, aggregated overall code edit pairs among student submissions:  
\[
\mathcal{L}_{\text{total}} = \lambda_{\text{C}} \mathcal{L}_{\text{C}} + \lambda_{\text{Rec}} \mathcal{L}_{\text{Rec}} + \lambda_{\text{Reg}} \mathcal{L}_{\text{Reg}},
\]
where \( \lambda_{\text{C}}, \lambda_{\text{Rec}}, \lambda_{\text{Reg}} \) are hyperparameters that strike a balance between the weights of each loss function.

\section{Experiments}\label{sec:exp}
In this section, we evaluate our proposed model on real student code submissions to assess its effectiveness in learning meaningful code-edit representations and generating personalized next-step code suggestions. We first describe the dataset and preprocessing steps, followed by details of our experimental setting and evaluation metrics. Next, we present the evaluation designs and baselines used to measure both code-edit representation quality and code generation performance. 
\subsection{Dataset}
We use the 2nd CSEDM Data Challenge dataset, hereafter referred to as the CSEDM dataset.\footnote{\url{https://sites.google.com/ncsu.edu/csedm-dc-2021/}. The dataset is called ``CodeWorkout data Spring 2019'' in Datashop (\url{pslcdatashop.web.cmu.edu}).} The dataset contains 246 students' 46,825 full submissions on a total of 50 programming questions in an introductory programming course. We use the version of the dataset that has accompanying test cases for 17 out of the 50 questions, released in \cite{duan2024test}, and run each submission against the individual test cases to obtain the test-case-mask of each submission. As a result, we obtain a total of around 8,000 combinations of $(A_1,A_2)$ and $(B_1,B_2)$ code-edit pairs, with balanced labels, i.e., $Y_{A,B}=1$ on around 50\% of such pairs. We split this dataset into three sets: Training, Validation, and Testing, across students: students in the Training set do not overlap with those in the Validation and Testing sets, preventing any code from leaking across the three sets. The individual data subsets contain roughly 90\%, 5\%, and 5\% of all students in the original dataset. 

\subsection{Experimental Settings}
We use the CodeT5-base model from Huggingface \cite{wolf2019huggingface} as our encoder-decoder model \cite{wang2021codet5}. We fine-tune it by minimizing our objective, using the AdamW optimizer with a learning rate of 1e-4, with a warmup for 10\% of steps and an effective batch size of 8 using gradient accumulation. We train for 100 epochs, which we find minimizes the loss on the validation set. Our best model has $\lambda_{C} = .5, \lambda_{Rec} = 2, \lambda_{Reg} = .5$.

We evaluate code generation using CodeBLEU \cite{ren2020codebleu}, which extends BLEU by incorporating syntax and semantics. It combines n-gram match, syntax tree match, data flow analysis, and weighted scoring to assess both structural correctness and functional equivalence. This ensures a more comprehensive evaluation of generated code beyond surface-level similarities. 

\subsection{Evaluation Design} \label{sec:evaluation_design}

\subsubsection{Edit-Embedding Based Next-Step Code Suggestion}
Our quantitative evaluation measures the effectiveness of our model-generated code-edit embeddings in producing next-step code suggestions for students. Given a student submission $A_1$, our goal is to generate an immediate next-step submission that helps address \emph{some} of their errors. Instead of providing a fully correct solution, we aim to offer a feasible path forward, guiding students toward incremental improvements. To achieve this, we incorporate a desired next-step test-case mask as input, ensuring that the generated code aligns with a predefined learning trajectory. The generated next-step code must satisfy the following two criteria:
\begin{itemize}
\item \textbf{Test-Case-Mask Adherence:} Generated codes should align with the desired test-case-mask, leveraging historical code-edit embeddings to ensure meaningful progression.
\item \textbf{Personalization:} The suggested code should retain structural and stylistic similarities to the student’s original submission, maintaining continuity in their problem-solving approach.
\end{itemize}

During the experiment, we simulate how our model would assist new students by using the test set as a proxy for unseen submissions. Our goal is to predict a plausible next-step code edit for a student, given their current code and a desired test-case outcome. Instead of generating an arbitrary next-step code, we leverage historical student edits to guide this prediction, ensuring that the suggestion follows realistic learning trajectories observed in past submissions.

For each test-set code pair \( (A_1, A_2) \) with corresponding test-case-masks $(T_A^1$, $T_A^2)$, we use \( A_1 \) as the starting code and \( T_A^2 \) as the desired test-case mask for the next-step code. To make a prediction, we retrieve past student submissions from the training set that followed a similar test-case transition \( (T_A^1, T_A^2) \). Let \( (H_1, H_2) \) be one such historical pair, with \( \mathbf{D}_H \) as its edit embedding. We then apply this edit transformation to the embedding of \( A_1 \), using \( \mathbf{e}(A_1) + \mathbf{D}_H \) as input to our model’s decoder to generate the predicted next-step code \( \hat{A}_2 \).  

To evaluate the quality of our generated code, we compare \( \hat{A}_2 \) with the actual student submission \( A_2 \) using CodeBLEU. A higher CodeBLEU score indicates that our model successfully predicts realistic next-step code. Additionally, to assess personalization, we compare \( \hat{A}_2 \) against both \( A_2 \) and \( H_2 \). If \( \hat{A}_2 \) is significantly closer to \( A_2 \) than \( H_2 \), it suggests that the generated next-step code is tailored to the original student's coding trajectory rather than being generic.  

As a final validation, we execute \( \hat{A}_2 \) on the input test cases and check whether it satisfies the desired test-case-mask \( T_A^2 \). This ensures that the generated code is not only personalized but also functionally relevant, effectively guiding students toward correcting their errors.\\\\
\textbf{Baseline: GPT-4o-based Next-Step Suggestion}
This baseline evaluates GPT-4o's ability to find a code that maintains the \textit{Test-Case-Mask Adherence} and \textit{Personalization} criteria. From the submission pairs $(A_1, A_2)$ in the test set, we provide the original code $A_1$, the problem description, the input test cases along with the desired test-case-mask $T_A^2$ and prompt GPT-4o to produce a code $\hat{A}_2$ that passes only test cases specified by $T_A^2$. \\\\
\textbf{Baseline: GPT-4o-based Decoding} This baseline evaluates GPT-4o's ability to represent and reproduce code changes through a two-step process. First, given a pair of code snippets—original code \( A_1 \) and modified code \( A_2 \)—we prompt GPT-4o to generate a structured textual description of the change, which serves as an \textit{edit representation}. Next, we provide GPT-4o with \( A_1 \) along with the generated edit representation and prompt it to produce the modified code \( \hat{A}_2 \).\\\\
\textbf{Baseline: CodeT5 for Decoding} In this baseline, we first use the average of the last hidden layer embeddings from the encoder to represent the code. Specifically, for two code snippets \( A_1 \) and \( A_2 \), we compute the embeddings \( \mathbf{e}(A_1) \) and \( \mathbf{e}(A_2) \) by averaging the last hidden layer outputs of the encoder for each code snippet. Next, the edit embedding \( \mathbf{D} \) is obtained by taking the difference between the two code embeddings:  \(\mathbf{D} = \mathbf{e}(A_2) - \mathbf{e}(A_1)\). This edit embedding is then passed through the decoder, which generates the modified code based on the learned representation of the code change. The generated code \( \hat{A}_2 \) is compared with \( A_2 \) using CodeBLEU to evaluate the quality of the code reproduction.

\section{Results and Discussion}\label{sec:results}
In this section, we show the results of our experiments to evaluate our model. We begin by analyzing the overall performance of our approach compared to baselines, followed by an ablation study to assess the contribution of individual components. Finally, we discuss qualitative insights and the implications of our findings.
\begin{table}[]
\centering
\caption{Average CodeBLEU score for personalizated next-step code suggestion using historical code-edit embeddings. }\label{tab:cb-history}
\begin{tabular}{l|ll|l|llll|}
\cline{2-8}
& \multicolumn{2}{c|}{\textbf{Baselines}}                                   & \multicolumn{1}{c|}{\multirow{2}{*}{\textbf{\begin{tabular}[c]{@{}c@{}}Ours \\ w/o ET\end{tabular}}}} & \multicolumn{4}{c|}{\textbf{Ours with Edit Transformation (ET)}}                                                                                                                                   \\ \cline{2-3} \cline{5-8} 
& \multicolumn{1}{c|}{\textbf{CodeT5}} & \multicolumn{1}{c|}{\textbf{GPT-4o}} & \multicolumn{1}{c|}{}     & \multicolumn{1}{c|}{\textbf{Full}}       & \multicolumn{1}{c|}{\textbf{$\lambda_{C} = 0$}} & \multicolumn{1}{c|}{\textbf{$\lambda_{Rec} = 0$}} & \multicolumn{1}{c|}{\textbf{$\lambda_{Reg} = 0$}} \\ \hline
\multicolumn{1}{|l|}{\textbf{All}}     & \multicolumn{1}{l|}{0.25, 0.25}      & 0.57, 0.54                         & \textbf{0.73, 0.54}                                                                                   & \multicolumn{1}{l|}{0.69, 0.54}          & \multicolumn{1}{l|}{0.69, 0.54}                 & \multicolumn{1}{l|}{0.14, 0.15}                   & \textit{0.72, 0.54}                               \\ \hline
\multicolumn{1}{|l|}{\textbf{If-else}} & \multicolumn{1}{l|}{0.25, 0.25}      & 0.63, 0.55                         & \textbf{0.79, 0.58}                                                                                   & \multicolumn{1}{l|}{\textit{0.75, 0.59}} & \multicolumn{1}{l|}{0.74, 0.59}                 & \multicolumn{1}{l|}{0.16, 0.17}                   & 0.73, 0.57                                        \\ \hline
\multicolumn{1}{|l|}{\textbf{String}}  & \multicolumn{1}{l|}{0.25, 0.25}      & 0.54, 0.42                         & 0.74, 0.39                                                                                            & \multicolumn{1}{l|}{\textit{0.75, 0.40}} & \multicolumn{1}{l|}{\textit{0.75, 0.40}}        & \multicolumn{1}{l|}{0.16, 0.20}                   & \textbf{0.83, 0.40}                               \\ \hline
\multicolumn{1}{|l|}{\textbf{Array}}   & \multicolumn{1}{l|}{0.25, 0.25}      & 0.52, 0.45                         & 0.65, 0.53                                                                                            & \multicolumn{1}{l|}{\textbf{0.72, 0.52}} & \multicolumn{1}{l|}{0.70, 0.52}                 & \multicolumn{1}{l|}{0.17, 0.17}                   & \textit{0.72, 0.58}                               \\ \hline
\end{tabular}
\end{table}
\subsection{Next-Step Code Suggestion Performance}


The next-step code suggestion experiment evaluates the effectiveness of our model in generating \textit{Personalized}, next-step codes that can fulfill the \textit{Test-Case-Mask Adherence} criteria using the code-edit embeddings from the historical submissions. This performance is shown in Tables \ref{tab:cb-history} and \ref{tab:stat-personalization}.

Table \ref{tab:cb-history} shows two values in each cell. The first value, \textbf{CBP}, is the average CodeBLEU score between our model generated next-step code $\hat{A}_2$ and the actual student-written next code $A_2$ in the test set. The second value, \textbf{CBH}, is the average CodeBLEU score between $\hat{A}_2$ and the retrieved code $H_2$ from the historical submissions associated with the desired \textit{Test-Case-Mask}. A higher CBP is our primary goal, while a lower CBH indicates good personalization. The first row includes results for all the problems, while subsequent rows are results on specific topics in the dataset. We see that both versions of our model outperform the GPT-4o baseline. The reason behind this performance is that our model learns code-edit that effectively summarize code changes across student submissions. GPT-4o, when asked to maintain the \textit{Test-Case-Mask Adherence} in the prompt, erroneously produces codes that passes all the test cases, hence lowering the CBP score. Therefore, the code produced by GPT-4o looks significantly different than the student's starting code $A_1$, which may confuse the student. On the contrary, our model produces codes that are close to the student's.

\begin{table}[h]
\centering
\caption{Statistics of execution for generated codes.}\label{tab:stat-personalization}
\begin{tabular}{l|l|l|}
\cline{2-3} & \textbf{GPT-4o} & \textbf{Ours} \\ 
\hline
\multicolumn{1}{|l|}{\textbf{Exact Match of Test-Case-Masks}}           & 2\%            & \textbf{56\%} \\ \hline
\multicolumn{1}{|l|}{\textbf{Passes All Test Cases}} & 79\%           & \textbf{5\%}  \\ \hline
\end{tabular}
\end{table}

Table \ref{tab:stat-personalization} presents the quantitative evaluation for the \textit{Test-Case-Mask Adherence} criterion. We evaluate generated code by executing it against the given test cases, computing the resulting test-case masks, and comparing them to the expected test-case mask. The first row of the table demonstrates that our model generates code that aligns significantly better with the desired test-case masks compared to GPT-4o. The second row measures how well models produce code that matches the intended test-case behavior. Even when our model does not perfectly adhere to the specified mask, its outputs remain close, rarely generating code that passes all test cases. In contrast, GPT-4o tends to produce fully correct solutions almost every time. However, we acknowledge that while GPT-4o consistently generates syntactically correct code, our model often produces minor syntactical errors in outputs, which remains a limitation.

\begin{table}[]
\centering
\caption{Average CodeBLEU score for code reconstruction from Edit Embeddings (EE) and Cross Transformation (CT).}\label{tab:cb-edit-cross}
\begin{tabular}{l|ll|l|llll|}
\cline{2-8}
& \multicolumn{2}{c|}{\textbf{Baselines}} & \multicolumn{1}{c|}{\multirow{2}{*}{\textbf{\begin{tabular}[c]{@{}c@{}}Ours \\ w/o ET\end{tabular}}}} & \multicolumn{4}{c|}{\textbf{Ours with Edit Transformation (ET)}} \\ \cline{2-3} \cline{5-8} 
& \multicolumn{1}{c|}{\textbf{CodeT5}} & \multicolumn{1}{c|}{\textbf{GPT-4o}} & \multicolumn{1}{c|}{}                                                                                 & \multicolumn{1}{c|}{\textbf{Full}} & \multicolumn{1}{c|}{\textbf{$\lambda_{C} = 0$}} & \multicolumn{1}{c|}{\textbf{$\lambda_{Rec} = 0$}} & \multicolumn{1}{c|}{\textbf{$\lambda_{Reg} = 0$}} \\ \hline
\multicolumn{1}{|l|}{\textbf{EE}}  & \multicolumn{1}{l|}{0.24}            & \textbf{0.90}                      & 0.75                                                                                                  & \multicolumn{1}{l|}{\textit{0.85}} & \multicolumn{1}{l|}{0.84}                       & \multicolumn{1}{l|}{0.19}                         & 0.84                                              \\ \hline
\multicolumn{1}{|l|}{\textbf{CT}} & \multicolumn{1}{l|}{0.25} & 0.81 & 0.77 & \multicolumn{1}{l|}{\textbf{0.87}} & \multicolumn{1}{l|}{0.74} & \multicolumn{1}{l|}{0.18} & \textit{0.86}                                     \\ \hline
\end{tabular}
\end{table}

Table \ref{tab:cb-edit-cross} shows the overall code reconstruction performance in two cases. The \textbf{Edit Embedding (EE)} case denotes the experiment where we used the code-edit embeddings from the test set, rather than finding a similar one from the historical database. In the \textbf{Cross Transformation (CT)} case, we used the code-edit embeddings of the submission pair in the same row of the test set if it's labeled as similar. In GPT-4o baseline, the result comes from the \textbf{GPT-4o-based Decoding} (section \ref{sec:evaluation_design}) experiment that uses an edit description produced by GPT-4o, and later uses the edit description along with the original code to generate the next step code. We see that GPT-4o performs better in EE, since the textual edit description of GPT-4o stores much more information than the dense edit-embeddings of our model. However, this edit description does not generalize as well as our edit embeddings, and we can see from the CT row of the table. Our model's edit embeddings are more generalizable in the more realistic setting than GPT-4o's edit descriptions.

\begin{figure}[!t]
    \centering
    \begin{subfigure}{0.49\textwidth}
        \centering
        \includegraphics[width=\linewidth]{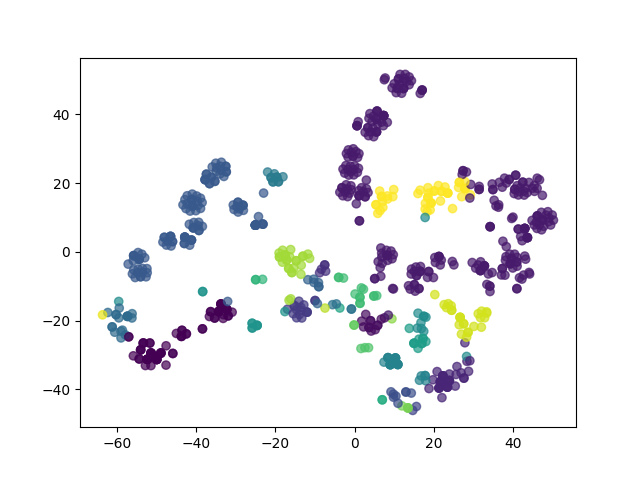}
        \caption{$\lambda_{Rec} = 0$}
    \end{subfigure}
    \begin{subfigure}{0.49\textwidth}
        \centering
        \includegraphics[width=\linewidth]{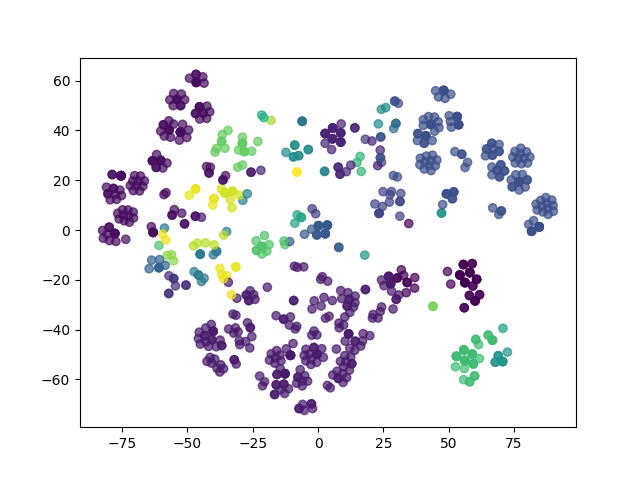}
        \caption{$\lambda_{Reg} = 0$}
    \end{subfigure}
    
    \begin{subfigure}{0.49\textwidth}
        \centering
        \includegraphics[width=\linewidth]{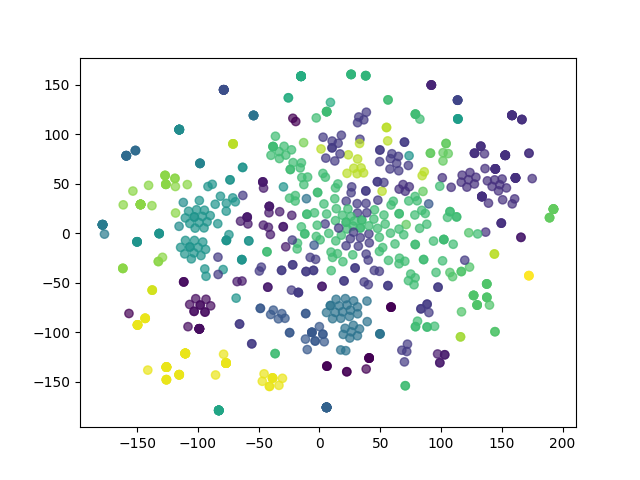}
        \caption{$\lambda_{C} = 0$}
    \end{subfigure}
    \begin{subfigure}{0.49\textwidth}
        \centering
        \includegraphics[width=\linewidth]{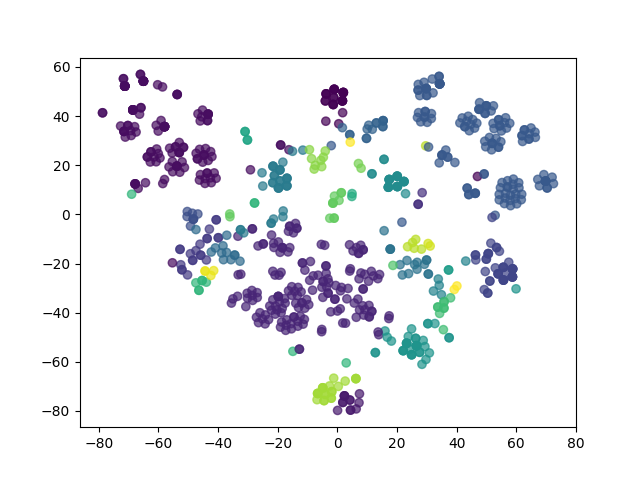}
        \caption{Full}
    \end{subfigure}
    \caption{Low dimensional visualization of edit embedding clusters in the test set produced by our model. Each figure denotes the output for different ablation settings. The same color of points belong in the same cluster.}
    \label{fig:clusters-test-set}
    \vspace{-.5cm}
\end{figure}

\subsection{Ablation Experiments}
The ablation experiments on $\lambda_{C}$, $\lambda_{Rec}$, and $\lambda_{Reg}$ offer valuable insights into their roles. Tables \ref{tab:cb-history} and \ref{tab:cb-edit-cross}, as well as Figure \ref{fig:clusters-test-set},  illustrate the impact of nullifying each $\lambda$. Nullifying $\lambda_{Rec}$ significantly degrades the reconstruction performance, since CodeT5 is not pre-trained to generate code from average token embeddings. However, it has little to no effect on producing well-formed clusters in the code-edit embedding space, as shown in Figure \ref{fig:clusters-test-set}(a). In contrast, both $\lambda_{C}$ and $\lambda_{Reg}$ provide moderate improvements in performance, highlighting their importance in shaping the vector space. Nullifying $\lambda_{Reg}$ leads to decreased performance in Table \ref{tab:cb-edit-cross}, but increases performance in Table \ref{tab:cb-history}. Its effect is negligible in Figure \ref{fig:clusters-test-set}(b), suggesting that its impact is inconclusive. The effect of nullifying $\lambda_{C}$ is more pronounced, as evidenced by the absence of visible clusters in Figure \ref{fig:clusters-test-set}(c). In reconstruction tasks, nullifying $\lambda_{C}$ reduces performance in some cases while leaving it unchanged in others.

\begin{table}[t]
\centering
\vspace{-.5cm}
\caption{GPT-4o generated summary of a cluster with off-by-one error.} \label{tab:cluster-summary-example}
\begin{tabular}{c|c}
\hline
\textbf{Initial Code} & \textbf{Next Code}\\ 
\hline
\begin{minipage}[t]{0.49\linewidth}
\begin{lstlisting}[breaklines, linewidth=\textwidth, basicstyle=\ttfamily]
public String zipZap(String str){
  String newStr = """";
  int length = str.length() - 1;
  for (int i = length; i >= 3; i--){
    if (str.charAt(i) == 'p' && str.charAt(i-2) == 'z'){
      newStr = str.substring(0, i-1) + str.substring(i);
    }
  }
  return newStr;
}
\end{lstlisting}

\end{minipage}     
&  
\begin{minipage}[t]{0.49\linewidth}
\begin{lstlisting}[breaklines, linewidth=\textwidth, basicstyle=\ttfamily]
public String zipZap(String str){
  String newStr = """";
  int length = str.length() - 1;
  for (int i = length; i >= 2; i--){
    if (str.charAt(i) == 'p' && str.charAt(i-2) == 'z'){
      newStr = str.substring(0, i-1) + str.substring(i);
    }
  }
  return newStr;
}
\end{lstlisting}
\end{minipage}\\
\hline
\begin{minipage}[t]{0.49\linewidth}
\begin{lstlisting}[breaklines, linewidth=\textwidth, basicstyle=\ttfamily]
public boolean xyBalance(String str){
  int length = str.length() - 1;
  int a = str.lastIndexOf(""x"");
  for (int i = a; i < length; i ++){
    if(str.charAt(i) == 'y'){
      return true;
    }
  }
  return false;
}
\end{lstlisting}
\end{minipage}     
&  
\begin{minipage}[t]{0.49\linewidth}
\begin{lstlisting}[breaklines, linewidth=\textwidth, basicstyle=\ttfamily]
public boolean xyBalance(String str){
  int length = str.length() - 2;
  int a = str.lastIndexOf(""x"");
  for (int i = a; i < length; i ++){
    if(str.charAt(i) == 'y'){
      return true;
    }
  }
  return false;
}
\end{lstlisting}
\end{minipage}\\
\hline
\begin{minipage}[t]{0.49\linewidth}
\begin{lstlisting}[breaklines, linewidth=\textwidth, basicstyle=\ttfamily]
public String getSandwich(String str){
  return str.substring(str.indexOf(""bread"")+ 5 + str.lastIndexOf(""bread""));    
}
\end{lstlisting}
\end{minipage}     
&  
\begin{minipage}[t]{0.49\linewidth}
\begin{lstlisting}[breaklines, linewidth=\textwidth, basicstyle=\ttfamily]
public String getSandwich(String str){
  return str.substring(str.indexOf(""bread"")+ 4 + str.lastIndexOf(""bread""));    
}
\end{lstlisting}
\end{minipage}\\
\hline
\end{tabular}
\vspace{5pt}
\begin{tabular}{c}
\textbf{Summary Generated by GPT-4o} \\ 
\hline
\begin{minipage}[t]{\linewidth}
The common error in the `Initial Code' across all pairs involves incorrect boundary conditions in loops or substring operations, often related to off-by-one errors. In the `zipZap' examples, the loop's starting condition is incorrect, starting from an index that is too high. Similarly, in `xyBalance', the loop runs to an incorrect boundary. In `getSandwich', the substring index calculation is off by one. The common debugging pattern used by the students to get to the `Next Code' involves adjusting these boundary conditions to the correct indices, typically by decreasing the length limit or correcting the index calculations by one.
\end{minipage} \\ \hline
\end{tabular}
\vspace{-.7cm}
\end{table}

\subsection{Debugging Behavior using Cluster Summaries}
We conduct an experiment to assess the quality of the clusters generated by our model. First, we compute code-edit embeddings across pairs of student code submissions over the entire dataset. We then apply K-means clustering \cite{jin2011k} to group these embeddings within the vector space. Finally, we prompt GPT-4o to analyze representative code-change instances from each cluster and generate concise summaries describing the nature of the changes observed.  

The GPT-4o-generated summaries offer valuable insights into common debugging behaviors and error patterns. Table \ref{tab:cluster-summary-example} shows one such example, where the clustering effectively identifies a recurring pattern of off-by-one errors across multiple code submissions. Off-by-one errors are among the most prevalent errors made by developers, including novices \cite{sellik2021learning}. In our example, these errors primarily stem from incorrect boundary conditions in loops or substring operations. Specifically, in the `zipZap' examples, the loop's starting condition is incorrect, while in `xyBalance,' the loop boundary is miscalculated. Similarly, in `getSandwich,' the substring index calculation is off by one. The debugging pattern observed in the `Next Code' column suggests that students commonly resolve these issues by adjusting boundary conditions, either by modifying the length limit or correcting index calculations.  

Despite the contrastive objective being based solely on problem-exclusive test-case-masks, our model still learns to cluster code-edit embeddings from different problems that correspond to similar debugging actions, which suggests that the learned embeddings capture meaningful structural similarities in code modifications beyond surface-level problem constraints. As a result, this clustering approach not only pinpoints specific areas of difficulty but also highlights recurrent debugging strategies across different problems. Identifying these common error patterns can facilitate the development of targeted feedback mechanisms, helping students recognize and address frequent errors more effectively.

\section{Conclusion and Future Works}
In this work, we explored the use of contrastive learning for generating meaningful vector representations of code changes, with the goal of identifying debugging patterns in student submissions. Our approach clusters code-edit embeddings based on test case transitions, revealing structured patterns of errors and their corresponding fixes. Our encoder-decoder model learns code-edit embeddings that capture both syntactic and semantic similarities, allowing for meaningful and gradual guidance in debugging by leveraging the historical edit embeddings. Unlike GPT-4o-based methods that often generate fully correct solutions, our approach ensures a balance between assistance and student effort, preserving the `Desirable Difficulty’ principle in learning. Additionally, by clustering edit embeddings, we identify common student errors and debugging patterns, offering insights that can enhance instructional strategies. Experimental results validate that our model significantly improves upon existing approaches, providing a more learner-centric feedback mechanism. 

In the future, we expect to improve the generation of syntactically correct code by our method. A post-processing step using LLM can be one way of solving this problem. We would also like to explore the utilization of the clusters in a pedagogically meaningful way. We will explore extending our methodology to more diverse programming languages and integrating it into real-world educational platforms for broader impact.

\bibliographystyle{plainnat}
\bibliography{article}

\end{document}